\documentclass[journal=jpclcd,manuscript=article]{achemso}
\usepackage[utf8]{inputenc}
\usepackage[T1]{fontenc}
\usepackage[version=3]{mhchem}
\usepackage{soul}
\usepackage{bm}
\usepackage{threeparttable}
\usepackage{color}


\begin{tocentry}
\includegraphics{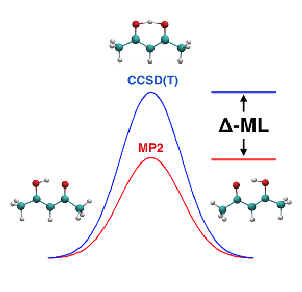}
\end{tocentry}

\author{Chen Qu}
\affiliation{Department of Chemistry \& Biochemistry, University of Maryland, College Park, Maryland 20742, U.S.A.}
\author{Paul L. Houston*}
\email{plh2@cornell.edu}
\affiliation{Department of Chemistry and Chemical Biology, Cornell University, Ithaca, New York
14853, U.S.A. and Department of Chemistry and Biochemistry, Georgia Institute of
Technology, Atlanta, Georgia 30332, U.S.A}
\author{Riccardo Conte*}
\email{riccardo.conte1@unimi.it}
\affiliation{Dipartimento di Chimica, Universit\`{a} degli Studi di Milano, via Golgi 19, 20133 Milano, Italy}
\author{Apurba Nandi}
\affiliation{Department of Chemistry and Cherry L. Emerson Center for Scientific Computation, Emory University, Atlanta, Georgia 30322, U.S.A.}
\author{Joel M. Bowman*}
\email{jmbowma@emory.edu}
\affiliation{Department of Chemistry and Cherry L. Emerson Center for Scientific Computation, Emory University, Atlanta, Georgia 30322, U.S.A.}

\title{Breaking the Coupled Cluster Barrier for Machine Learned Potentials of Large Molecules: The Case of 15-atom Acetylacetone}

\begin{document}


\begin{abstract}
Machine-learned potential energy surfaces (PESs) for molecules with more than 10 atoms are typically forced to use lower-level electronic structure methods such as density functional theory and second-order M{\o}ller-Plesset perturbation theory (MP2). While these are efficient and realistic, they fall short of the accuracy of the ``gold standard'' coupled-cluster method, especially with respect to reaction and isomerization barriers. 
We report a major step forward in applying a $\Delta$-machine learning method to the challenging case of acetylacetone, whose MP2 barrier height for H-atom transfer is low by roughly 1.5 kcal/mol relative to the benchmark CCSD(T) barrier of 3.2 kcal/mol. From a database of 2151 local CCSD(T) energies, and training with as few as 430 energies, we obtain a new PES with a barrier of 3.49 kcal/mol  in agreement with the LCCSD(T) one of 3.54 kcal/mol and close to the benchmark value. 
Tunneling splittings due to H-atom transfer are calculated using this new PES, 
providing improved estimates over previous ones obtained using an MP2-based PES.
\end{abstract}

\flushbottom
\maketitle

\thispagestyle{empty}

There has been dramatic progress in using regression methods from machine learning (ML) to develop potential energy surfaces (PESs) for systems with more than five atoms, based on fitting thousands of CCSD(T) energies.\cite{bowman11, ARPC2018, Fu18, guo20} However, the CCSD(T) method, because it scales as $N^7$, where $N$ is the system size, is too computationally demanding for PES fits of systems with more than 10 heavy atoms. (This number of atoms is rightly not considered a ``large molecule'' by many readers; however, it is used here as a computational boundary for the CCSD(T) method.)   One 10-atom PES using the method we are aware of is the formic acid dimer \ce{(HCOOH)2},\cite{Qu2016} which contains 8 heavy atoms. This was a major computational effort at the CCSD(T)-F12a/haTZ (VTZ for H and aVTZ for C and O) level of theory. This PES, which does not dissociate, was obtained with only 13 475 energies. A 9-atom PES for the chemical reaction Cl+\ce{C2H6} was recently reported using a composite MP2/CCSD(T) method.\cite{cazko9atom}   Both of these PESs were fit using Permutationally Invariant Polynomial (PIP) regression. Examples of potentials for 6 and 7-atom chemical reactions which are fits to tens of thousands or even hundred thousand CCSD(T) energies have also been reported.\cite{bowman11, Fu18, guo20, Furoam20, HCH3OH}

The 10-atom CCSD(T)-barrier is due both to the steep scaling with $N$ and the increasing dimensionality of the PES, which requires larger data sets. Thus, given the intense interest, and progress, in moving to larger molecules and clusters, where high-level methods are prohibitively expensive, the use of lower-level methods such as Density Functional Theory (DFT) and second-order M{\o}ller-Plesset perturbation (MP2) theory is understandable.  These methods also provide analytical gradients and this is an important source of data needed for larger systems. Our group has made use of this approach for PIP PESs of $N$-methyl acetamide,\cite{QuBowman2019, NandiBowman2019} glycine\cite{conte_glycine20} and tropolone.\cite{tropolone20} 

We also recently reported PIP and fragmented PIP PESs (see below for some details) for 15-atom acetylacetone (AcAc),\cite{QuPCCP2020} using MP2 energies and gradients from Meuwly and co-workers,\cite{meuwly20} supplemented by us with roughly 500 additional configurations. This PES has a barrier for symmetric H-atom transfer of 2.13 kcal/mol (745 cm$^{-1}$) in close agreement with the direct MP2 value of 2.18 kcal/mol (763 cm$^{-1}$). However, that value of the barrier is more than 1 kcal/mol below the reported CCSD(T)/aug-cc-pVTZ one of 3.2 kcal/mol.\cite{acac15}  It is expected that this error in the MP2-based PES leads to a large overestimate of the tunneling splitting for the ground vibrational state H-atom transfer.  Nevertheless, the splitting was obtained with the MP2-based PES, using full-dimensional diffusion Monte Carlo calculations. The splitting is 160 cm$^{-1}$ with an uncertainty of 15 cm$^{-1}$. Using a simple 1d model, a splitting of 113 cm$^{-1}$ was obtained for a barrier of 2.2 kcal/mol and 74 cm$^{-1}$ for scaled barrier of 3.2 kcal/mol. This simple 1d estimate for the larger barrier is not expected to be quantitative; however, it does confirm that a large decrease in the splitting with increasing the barrier by 1 kcal/mol can be expected. 

This magnitude of the error in chemical barriers is typical for MP2 and DFT accuracy, compared to benchmark  CCSD(T) results.  In general, MP2 and DFT geometries and harmonic frequencies are relatively more accurate than barrier heights.  Thus, there is a strong motivation to improve a PES based on a lower-level method such as DFT and MP2 when the focus is on barriers. 

Recent approaches to do this, using ML, aim to bring a PES based on a low-level of electronic theory to a higher level.  There are two approaches currently being investigated to accomplish this goal.  One is transfer learning (TL), which has been developed extensively in the context of artificial neural networks,\cite{TL_ieee} and much of the work in that field has been brought into chemistry.\cite{roit19, Tkatch2018, Tkatch19, Stohr2020, meuwly20}
The basic idea of TL is that a fit obtained from one source of data (perhaps a large one) can be corrected for a related problem by using limited data and by making hopefully small training alterations to the parameters obtained in the first fit. Therefore, in the present context of PES fitting, an ML-PES fit to low-level electronic energies/gradients can be reused as the starting point of the model for an ML-PES with the accuracy of a high-level electronic structure theory. As noted, this is typically done with artificial neural networks, where weights and biases trained on lower-level data hopefully require minor changes in response to additional training using high-level data. 

The other approach is $\Delta$-machine learning.  In this approach a correction is made to a property obtained using an efficient, low-level \textit{ab initio} theory.\cite{Lilienfeld15, Tkatch19, Tkatch2018, Stohr2020, meuwly20} The focus of most work on TL or $\Delta$-learning  has been on developing transferable force fields, with applications mainly in the thermochemistry and molecular dynamics simulations. 

Meuwly and co-workers applied TL using thousands of local CCSD(T) energies to improve their MP2-based neural network PESs for malonaldehyde, acetoacetaldehyde and acetylacetone.\cite{meuwly20} We recently proposed and tested a $\Delta$-learning approach, that uses a small number of CCSD(T) energies, to correct a PES based on DFT electronic energies and gradients.\cite{deltaML2021} The method was validated for PESs of small molecules, \ce{CH4} and \ce{H3O+}, and for 12-atom $N$-methyl acetamide. In all cases, the coupled cluster energies were obtained over the same large span of configurations used to get the lower-level PES.  For $N$-methyl acetamide these included the $cis$ and $trans$ isomers and the saddle points separating them. 

Here we apply this $\Delta$-learning approach to 15-atom AcAc, \ce{C5H8O2}.  The approach is to construct a high-level, coupled-cluster-level PES starting from a lower level MP2, by using a correction PES. This correction PES is a fit to a small number of high-level \textit{ab initio} energies that spans the same range of configurations used to obtain the lower-level PES. Explicitly, the corrected high-level PES, denoted $V_{LL\rightarrow CC}$, is given by
\begin{equation}
   V_{LL\rightarrow CC} = V_{LL} + \Delta V_{CC-LL}, 
\end{equation}
where $V_{LL}$ is the lower-level PES and $\Delta V_{CC-LL}$ is the correction PES.  In the present application to AcAc we calculated 2151 LCCSD(T)-F12/cc-pVTZ-F12 energies\cite{LCCSDT} and performed training on subsets of these ranging in size from 430 to 1935. By contrast, $V_{LL}$ was fit using a data size of 250 884 MP2 energies and gradients.

\hspace{\parindent}In the PIP approach to fitting,\cite{Braams09} the potential $V$ is represented in compact notation  by 
\begin{equation}
V(\bm{x})= \sum_{i=1}^{n_p} c_i p_i(\bm{x}),
\label{eq1}
\end{equation}
where $c_i$ are linear coefficients, $p_i$ are PIPs, $n_p$ is the total number of polynomials for a given maximum polynomial order and $\bm{x}$ are Morse variables.  For example, $x_{\alpha \beta}$ is given by exp($-r_{\alpha \beta}/\lambda$), where $r_{\alpha \beta}$ is the internuclear distance between atoms $\alpha$ and $\beta$. The range (hyper) parameter, $\lambda$, was chosen to be 2 bohr. The linear coefficients are obtained using standard least squares methods for a large data sets of electronic energies (and for large molecules gradients as well) at scattered geometries. 

For molecules of more than ten atoms, the size of the PIP basis can become a computational bottleneck. This size depends in a complicated and non-linear way with respect to the maximum polynomial order, the number of Morse variables, and the order of the symmetric group.\cite{Braams09} While we have been able to use a full PIP basis even for 15-atom tropolone\cite{tropolone20} and AcAc\cite{QuPCCP2020}, we have shown that the fragmented PIP, which can be applied to larger molecules, performs very well and runs faster than the full PIP basis. 

The fragmented PIP basis is obtained by fragmenting a molecule into groups of atoms.  A PIP basis for each group can be calculated rapidly and then combined with those of other groups to provide a compact and still precise representation of the PES.\cite{QuBowman2019} Indeed this has been verified for $N$-methyl acetamide, \cite{QuBowman2019, NandiBowman2019}, tropolone,\cite{tropolone20} and AcAc.\cite{QuPCCP2020} Note that these PESs use the most recent software that include gradients in the fit and produces gradients on output.\cite{NandiQuBowman2019, conte20, msachen}

\begin{figure}[htbp!]
\begin{center}
\includegraphics[width=0.6\textwidth]{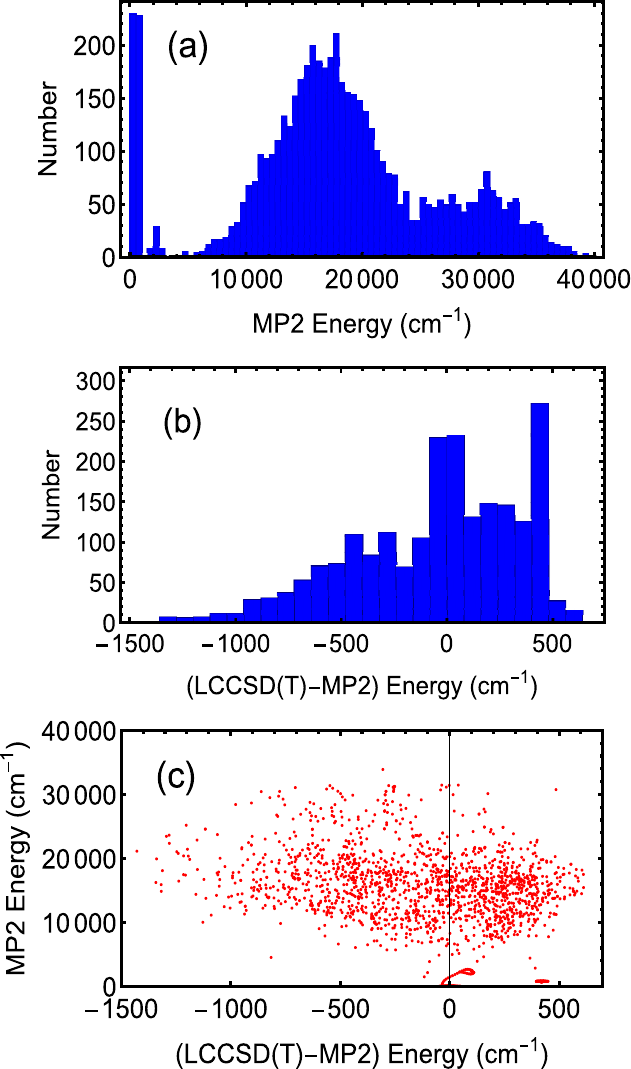}
\end{center}
\caption{Energies of MP2 and difference fits. (a) Histogram of MP2 energies for the MP2-based PES. The bin size is 500 cm$^{-1}$. (b) Histogram of difference between the LCCSD(T) and MP2 energies used in the $\Delta V_{CC-LL}$ fit. The bin size is 80 cm$^{-1}$.  (c) Correlation between MP2 energies in (a) and the LCCSD(T)-MP2 energies in (b).}
\label{fig: EHist}
\end{figure}

The histogram of the MP2 data base from our AcAc PES\cite{QuPCCP2020} is shown in Fig. \ref{fig: EHist}(a); it includes 5454 geometries.  The MP2 energies are relative to the MP2 global minimum.  The two lowest bars of the histogram are primarily the 454 points added by our group via an AIMD trajectory or random grids starting from or centered on the global minimum (GM) and H-atom transfer transition-state saddle point geometries but not specifically chosen to be along the reaction coordinate. Hereafter, unless indicated otherwise, the H-atom transfer saddle point is denoted ``SP". The AIMD trajectory was run at 400 cm$^{-1}$ for 200 steps starting from the GM, and 50 geometries/energies were used. The grids were made from taking the Cartesian coordinates of the GM and SP and  varying, for each geometry, all of the 39 coordinates by random number between the coordinate $\pm~ del$. For each of these starting points, approximately 75 geometries were generated with $del$ = 0.001 \AA,  and approximately 150 geometries were generated with $del$ = 0.004 \AA. In all, our group calculated 454 points for the MP2 data base. The majority of of the geometries and energies (5000 points) were kindly provided to us by the Meuwly group based on their AIMD trajectories.

The data base we used to calculate the difference potential, $\Delta V_{CC-LL}$, is described by the energy histogram in Fig. \ref{fig: EHist}(b). The LCCSD(T) energies are relative to the LCCSD(T) energy at the MP2 global minimum. The data base is composed of the 500 geometries that are clustered near to the GM, as determined by choosing the RMS bond difference for the 105 bond lengths to be less than 0.3 \AA~ when comparing the geometry to that of the GM, and another 1651 taken by a) choosing a random integer that indicated their position on a list of the remaining MP2 points, b) discarding choices whose MP2 energies were more than 30 000 cm$^{-1}$, and c) stopping the selection when the requisite number of choices had been made. Again, the aim was to produce a difference PES that spans the same range of configurations as the original lower-level PES, namely, one that included points near the GM and SP but not specifically chosen to be along the reaction coordinate. The difference data base shows a peak near zero energy difference with a distribution stretching from approximately -1500 to 800 cm$^{-1}$. 

Figure \ref{fig: EHist}(c) shows the correlation between the MP2 energies and those of the difference potential (LCCSD(T)-MP2). Note that the MP2 energies at which LCCSD(T) energies were calculated span the range from 0 to about 30 000 cm$^{-1}$.  It is clear that there is no systematic correlation between the MP2 energy and the difference energy. The largest energy differences are for geometries whose MP2 energies are between 10 000 and 20 000 cm$^{-1}$ whereas small differences are seen for MP2 energies as large as 30 000 cm$^{-1}$. The scatter of energies makes it clear that the difference potential is global in nature. Further, as can be seen, the density of points for the difference potential mirrors the density of MP2 energies. This is already an indicator the difference potential spans a very similar range of configurations as the original MP2 energies.  Further support comes from looking at the distribution of geometries discussed next.


\begin{figure}[htbp!]
\begin{center}
\includegraphics[width=0.5\textwidth]{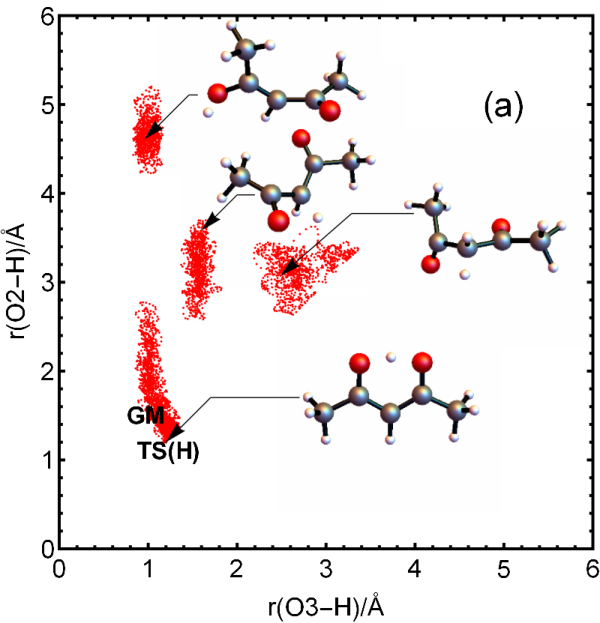}
\includegraphics[width=0.5\textwidth]{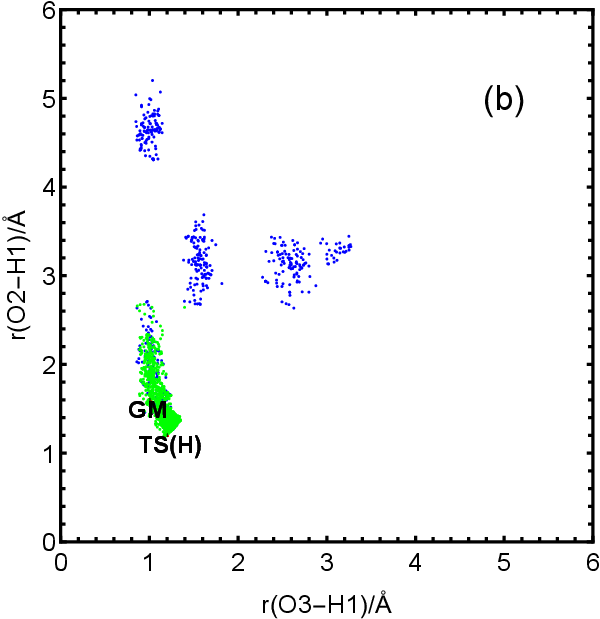}
\end{center}
\caption{(a) MP2 data base points showing representative structures. (b) LCCSD(T) data base points using all points near the GM (green) and other random scattered points (blue), as explained in the text. TS(H) denotes the H-transfer saddle point, which in the text is referred to as ``SP''.}
\label{fig: ScatterPlotsb}
\end{figure}

Fig. \ref{fig: ScatterPlotsb} shows that the distribution of geometries where LCCSD(T) calculations were performed overlaps the distribution where MP2 ones were performed, as it must in order to construct a difference potential. The plot axes show the two OH distances, where H is the transferring H atom. The numbering of atoms is shown in the Supporting Information (SI), and, because the distribution is permutationally symmetric, we have shown only the upper half by taking O2-H1 to be larger than O3-H1. Here, H1 is the hydrogen atom transferred and O2 and O3 are the symmetric oxygens to which it can be bonded. Note that other distances are changing as well, and this is indicated schematically by the structures shown in the figure. As seen, the LCCSD(T) set is sparse and dispersed. Note also that the clustering observed in both panels is due to the selection procedures adopted by the Meuwly group in the construction of their data set.

In addition to the LCCSD(T)-F12 energies calculated for the PES data base, we performed two benchmark calculations at the global minimum and at the saddle point for H-atom transfer. These two calculations found the optimum geometries and energies and determined as well the harmonic vibrational frequencies and normal coordinates.  While the LCCSD(T)-F12 calculations for just the energy at a single geometry took approximately 30 minutes using 12 cores of the 2.4 GHz Intel Xeon processors, the full optimization and frequency calculations took on the order of 73 days using the same number of processors.  This computational cost certainly underscores the infeasibility of doing even LCCSD(T) calculations for an AcAc PES.

The fit to the difference potential has to take account of the small data set, i.e., a maximum of 2151 energies. Therefore, the number of terms in the PIP basis has to be significantly less than this number to avoid overfitting. With this is mind, we used a PIP basis of maximum polynomial order of 2 and a symmetry designation of \{1,2,5,7\}.
The numbering scheme for this fit is shown in Fig. S1 of the SI, along with an explanation of the meaning of the symmetry designation. This basis contains just 85 PIPs and thus 85 linear coefficients to be determined by standard least-squares regression. 
This is the smallest PIP we have ever used, and the bonus is that we can examine small training data sets without concerns about overfitting.  Further, the evaluation of such a small PIP basis is fast.

We do note that in order to test the performance of the PES on configurations outside the training set, the largest training set consists of only 90\% of the total 2151 LCCSD(T) data points ($N=1935$ shown in Table \ref{tab:training_size}); the remaining 10\% (every 10$^{th}$ point in the full data set of 2151 points) is reserved for testing purpose. Three smaller data sets are also used to fit the PES: for $N=1075, 717, 430$, every $k^{th}$ point among the 2151 ones is picked as training points, where $k=2,3,4$ for $N=1075,717,430$, respectively, while the remaining points not included in the training set are for testing purpose.



Numerous metrics of the performance of the $\Delta$-ML approach are given in Table \ref{tab:training_size}.  Beginning with training and testing errors of the $\Delta V_{CC-LL}$ PESs using training sets of different sizes, we note again that the testing set consists of points not used for training, and the testing error is on the differences between LCCSD(T) and MP2 electronic energies. It can be seen that the testing error increases monotonically as the number of training points decreases, due to smaller coverage by the training data, as expected. However, this increase in the testing error is relatively small.

\begin{table}[htbp!]
\centering

\caption{Indicated RMS errors of the $\Delta V_{CC-LL}$ PESs using training sets of different sizes ($N$). Training and testing RMSs refer to energies, the barrier height is for symmetric H-transfer, the RMS errors in internuclear distances are given for the global minimum (GM) and H-atom transfer saddle point (TS(H)-SP). The mean absolute errors (MAE) are given for harmonic frequencies relative to benchmark LCCSD(T) results. Energies and frequencies are in cm$^{-1}$, and distances in \AA.}

\begin{tabular*}{\textwidth}{@{\extracolsep{\fill}} l c c c c c}
\hline
\hline\noalign{\smallskip}
 & $N$=1935 & $N$=1075 & $N$=717~ & $N$=430~ & 
   V$_{LL}$~ \\
\noalign{\smallskip}\hline\noalign{\smallskip}
 Training RMS   & ~99.6  & ~94.7  & ~98.8  & ~79.3  &   -    \\
 Testing RMS    & 107.5  & 113.7  & 123.0  & 155.9  &   -    \\
 Barrier height &  1218  &  1217  &  1219  &  1219  &  ~745  \\
 RMS (GM Geom)  & 0.0081 & 0.0076 & 0.0078 & 0.0070 & 0.0115 \\
 RMS (TS(H) SP Geom)  & 0.0041 & 0.0041 & 0.0041 & 0.0036 & 0.0026 \\
 MAE (GM freq)  &  12.2  &  12.1  &  13.3  &  13.7  &  17.3  \\
 MAE (TS(H) SP freq)  &  24.7  &  24.8  &  26.2  &  25.9  &  35.6  \\
\noalign{\smallskip}\hline
\hline
\end{tabular*}

\label{tab:training_size}
\end{table}

Next we consider the equilibrium geometries and normal mode frequencies of both the global minimum and the saddle point to H transfer, as well as the barrier height.
For geometries, we computed the root-mean-square (RMS) difference between the 105 internuclear distances from the PES and direct LCCSD(T)-F12 optimized geometries.
For harmonic frequencies, we calculated the mean absolute error (MAE) by comparing frequencies from PESs with direct LCCSD(T)-F12 ones.
All these are listed in Table \ref{tab:training_size} for four PESs with different training sets as well as for the low-level MP2 PES. We get excellent agreement for the geometries: in all four $\Delta$-ML PESs, the RMS differences of the 105 internuclear distances between PES and direct LCCSD(T) geometries are around 0.008 and 0.004 \AA~ for GM and SP, respectively, while the RMS differences between V$_{LL}$ and LCCSD(T) geometries are 0.0115 and 0.0026 \AA~for GM and SP. Therefore the geometry of GM is slightly improved using the $\Delta$-ML approach, and the SP geometry is still in good agreement with the LCCSD(T)-F12 one despite slightly increased RMS difference (only 0.0014 \AA). A plot of V$_{LL\rightarrow CC}$ inter-nuclear distances vs direct \textit{ab initio} ones is shown in Fig. S2 of the SI. It is perhaps worth noting that the largest distances (nearly 7 \AA) are between the H atoms in the two methyl rotors. 

The barrier height of the H-transfer motion on all 4 corrected PESs, each based on a different training set, is around 1218 cm$^{-1}$ (3.49 kcal/mol), in excellent agreement with the direct LCCSD(T)-F12 value of 1234 cm$^{-1}$ (3.54 kcal/mol). The best estimate of this barrier height is 1148 cm$^{-1}$ (3.29 kcal/mol), based on CCSD(T)-F12/aug-cc-pVTZ single point calculations at the LCCSD(T)-optimized geometries. So the $\Delta$-ML PES slightly overestimates the barrier height; nevertheless, it is a significant improvement over the MP2-based PES\cite{QuPCCP2020}, which has a barrier height of 745 cm$^{-1}$ (2.13 kcal/mol). 


The energies of 7 low-lying stationary points (including the GM and SP) are shown in Table~\ref{tab:energies}. TS(T)-I/II/III are three transition state saddle points with respect to the torsion of the two methyl rotors, and TS(HT)-I/II are two higher-order saddle points with imaginary frequencies in both H-transfer motion and the methyl torsion.
In nearly all cases, the energies of the stationary points are better captured by the V$_{LL\rightarrow CC}$ PESs than by the V$_{LL}$ one.

\begin{table}[ht]
\caption{Energies (in cm$^{-1}$) of the 7 stationary points, relative to the global minimum (GM), using indicated methods. The numbers in parentheses for the two $\Delta$-ML PESs refer to the size of training data.}

\begin{threeparttable}
\begin{tabular*}{\textwidth}{@{\extracolsep{\fill}} l r r r r}
\hline
\hline\noalign{\smallskip}
Stationary points & LCCSD(T) & V$_{LL\rightarrow CC}$ (1935) & V$_{LL\rightarrow CC}$ (430) & ~~~V$_{LL}$~~~ \\
\noalign{\smallskip}\hline\noalign{\smallskip}
   GM     &        0~~~~    &    0~~~ &    0~~~ &   0~~~ \\
   TS(H)-SP     &     1234~~~~    & 1218~~~ & 1219~~~ & 745~~~ \\
TS(T)-I   & 123 \tnote{a}~  &  165~~~ &  154~~~ & 160~~~ \\
TS(T)-II  & 488 \tnote{a}~  &  477~~~ &  481~~~ & 399~~~ \\
TS(T)-III & 581 \tnote{a}~  &  627~~~ &  623~~~ & 541~~~ \\
TS(HT)-I  & 1434 \tnote{a}~ & 1299~~~ & 1306~~~ & 820~~~ \\
TS(HT)-II & 1645 \tnote{a}~ & 1359~~~ & 1374~~~ & 864~~~ \\
\noalign{\smallskip}\hline
\hline
\end{tabular*}

\begin{tablenotes}
\item[a] LCCSD(T)-F12 calculations at MP2-optimized geometries
\end{tablenotes}
\end{threeparttable}

\label{tab:energies}
\end{table}

The harmonic frequencies of the global minimum and H-transfer saddle point from the MP2 PES (V$_{LL}$), the corrected PES (V$_{LL\rightarrow CC}$) using 1935 training points, and direct LCCSD(T)-F12 calculations are listed in Table S1 of the SI. For most of the modes, the differences between V$_{LL}$ and V$_{LL\rightarrow CC}$ frequencies are small, but for mode 32 of GM (OH stretch) and the imaginary-frequency mode of the H-transfer SP, the improvement of the $\Delta$-ML PES is significant. Again, the 4 $\Delta$-ML PESs based on different training sets achieved similar MAE in frequencies (around 13 cm$^{-1}$ for GM and 25 cm$^{-1}$ for the H-transfer SP, see Table \ref{tab:training_size}), and that is a significant improvement over the low-level PES, which has MAEs of 17.3 and 35.6 cm$^{-1}$ for GM and the H-transfer SP, respectively.

These results show that the $\Delta$-ML approach indeed improves the PES and brings it closer to coupled-cluster level of accuracy. This approach significantly improves the barrier height of H transfer, moderately improves the harmonic frequencies of GM and H-transfer SP, and slightly improves the optimized geometries of GM. More importantly, even with a training set as small as 430 points, the corresponding $V_{LL\rightarrow CC}$ PES is almost as good as the one fitted to 1935 points. Nevertheless, the fit using 1935 points is still our best one in terms of coverage of configurations and testing error, and so the results for zero-point energy and H-transfer tunneling splitting shown next are based on this fit.

Diffusion Monte Carlo (DMC) calculations were employed to compute the ground-state tunneling splitting of AcAc. Specifically, the simple unbiased algorithm, \cite{Anderson1975, Schulten} was used to calculate the ground-state energy, while DMC with fixed-node approximation\cite{Anderson1976} was used to calculate the energy of the excited state with respect to the H-transfer motion.

In the simple unbiased algorithm we use, an ensemble of random walkers is used to represent the nuclear wavefunction of the molecule. At each step, a random displacement in each degree of freedom is assigned to each walker, and this walker may remain alive (and may give birth to a new walker) or be killed by comparing its potential energy, $E_i$, with a reference energy, $E_r$. For the ground state, the probability of birth or death is given as:
\begin{align}
    & P_\text{birth} = \exp \left[ -(E_i - E_r)\Delta \tau\right] - 1 \ (E_i < E_r)\\
    & P_\text{death} = 1 - \exp \left[ -(E_i - E_r)\Delta \tau\right] \ (E_i > E_r),
\end{align}
where $\Delta \tau$ is the step size in imaginary time.
In fixed-node approximation for excited states, in addition to the process described above, any walker that crosses a node is instantly killed. In most cases the node is unknown in Cartesian coordinates, but for certain modes such as H-transfer in symmetric double-well, a very reasonable approximation can be made for the node as described in detail below.

After removing all dead walkers, the reference energy is updated using the equation
\begin{equation}
  E_r(\tau) = \langle V(\tau) \rangle - \alpha \frac{N(\tau) - N(0)}{N(0)},
\end{equation}
where $\tau$ is the imaginary time; $\langle V(\tau) \rangle$ is the average potential over all the walkers that are alive; $N(\tau)$ is the number of live walkers at time $\tau$; $\alpha$ is a parameter that can control the fluctuations in the number of walkers and the reference energy. Finally, the average of the reference energy over the imaginary time gives an estimate of ZPE (or the energy of the excited state in a fixed-node calculation).

For AcAc, DMC calculations were performed in Cartesian coordinates in full dimensionality. For fixed-node calculation, we assume that the node is $r_\text{H1O2}=r_\text{H1O3}$ (using the numbering scheme shown in Fig. S1 of the SI). Initially, the H1 atom is closer to one O, say O3, so if $r_\text{H1O3}$ of a walker becomes larger than $r_\text{H1O2}$, that walker crosses the node and would be instantly removed. An additional correction was made for the excited state by taking recrossing into consideration.\cite{Anderson1976}

Ten DMC simulations were performed for each state, and in each simulation, 30 000 walkers were equilibrated for 5000 steps, and then were propagated for 50 000 steps to compute the energy, with a step size of 5.0 au. In these simulations, $\sim 10^{10}$ potential energy evaluations are required; clearly these cannot be done without an efficient PES.

The zero-point energy of the corrected PES using 10 DMC calculations is $26 741 \pm 7$ cm$^{-1}$, while the energy of the excited state for the H-transfer motion from 10 fixed-node DMC calculations is $26 773 \pm 10$ cm$^{-1}$; therefore the tunneling splitting is 32 cm$^{-1}$ with an uncertainty of roughly $\pm 10$ cm$^{-1}$. As a comparison, the splitting using the MP2-based PES (i.e., V$_{LL}$) is 160 cm$^{-1}$. Such a significant decrease in the tunneling splitting is expected because the barrier height of V$_{LL \rightarrow CC}$, 1218 cm$^{-1}$, is significantly higher than the barrier height from V$_{LL}$, 745 cm$^{-1}$.

\begin{figure}[htbp!]
    \centering
    \includegraphics[width=0.7\textwidth]{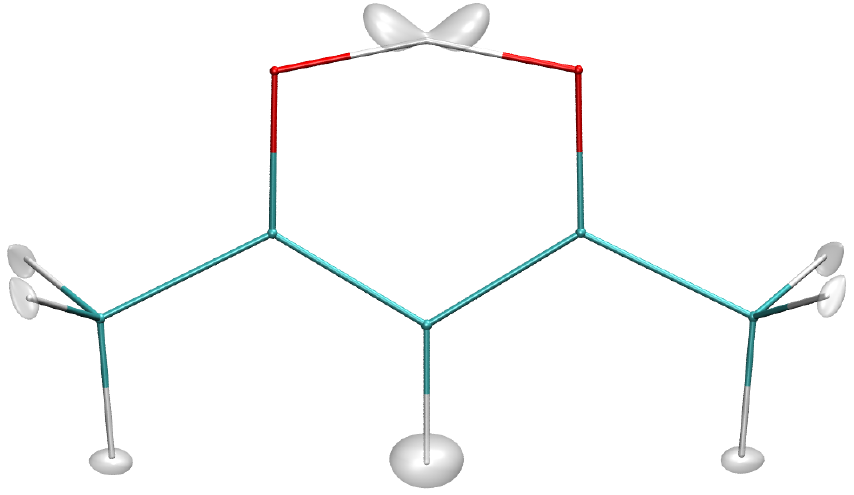}
    \caption{The ground state wavefunction of hydrogen from the DMC calculations.}
    \label{fig:wavefun}
\end{figure}

The ground state wavefunction from DMC calculations using the new PES is shown in Figure \ref{fig:wavefun}. It can be seen that the largest magnitude of the transferring H atom is closer to one of the O atoms, which is a character of the GM. However, the orientations of the two methyl rotors are the same as the H-transfer SP. Based on the DMC wavefunction, we cannot draw a definitive conclusion whether the ground state of AcAc has a $C_s$ or $C_{2v}$ structure. In fact, the ground state structure of AcAc is under debate; Meuwly et al. finds a slight preference of the $C_s$ structure by inclusion of MP2 ZPE difference to the CCSD(T) electronic energies of GM and H-transfer SP,\cite{acac15} while the microwave experiments by Caminati and Grabow support a $C_{2v}$ structure.\cite{Caminati2006} 

We also performed DMC calculations for the singly deuterated isotopologue of AcAc, but the energies of the ground state and excited state are too close so that the splitting is smaller than the uncertainty in the DMC calculations. Therefore, we could not obtain a reliable estimate of the tunneling splitting for the deuterated AcAc using DMC.   

\begin{figure}[htbp!]
    \centering
    \includegraphics[width=0.7\textwidth]{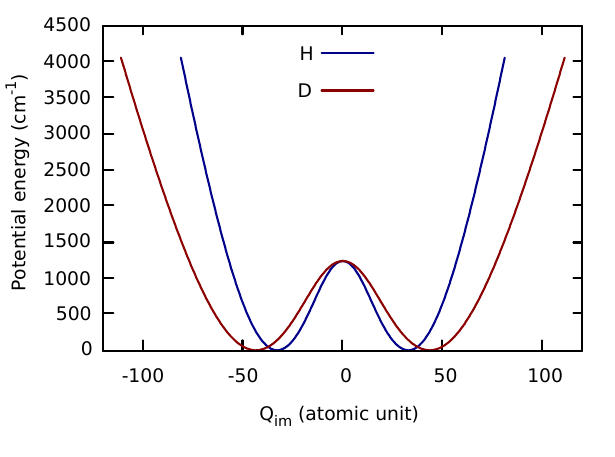}
    \caption{One-dimensional $V(Q_{im})$ path for H and D transfer in AcAc. The barrier heights have been ``morphed'' to agree with LCCSD(T)-F12 value.}
    \label{fig:qim}
\end{figure}

We also applied an approximate 1d approach to obtain the tunneling splittings. This method has been described in detail previously.\cite{qim2008}
It was used in our work on AcAc based on the MP2 PES (V$_{LL}$).\cite{QuPCCP2020}
Briefly, a 1d potential, denoted $V(Q_{im})$, which is the minimum energy path as a function of the imaginary-frequency mode ($Q_{im}$) of the H-transfer saddle point, was obtained by optimizing all the other coordinates at fixed $Q_{im}$ values using the $V_{LL\rightarrow CC}$ PES except the methyl rotors, which cannot be described using rectilinear normal coordinates. These are held fixed at the saddle point values all the way along the path. Due to fixed methyl orientation and fitting error, the barrier height of this 1d $Q_{im}$ path is 1055 cm$^{-1}$ and it is 179 cm$^{-1}$ lower than the LCCSD(T) value. Therefore, we ``morphed'' this 1d potential using the same strategy as described previously \cite{QuPCCP2020} so that it gives the correct barrier height (1234 cm$^{-1}$). The two mass-scaled 1d potentials employed (one for H and the other for D-transfer) are shown in Fig. \ref{fig:qim}.

The splittings are obtained simply using 1d-DVR calculations \cite{dvr1992} of the energies of the ground and first excited states on the morphed $V(Q_{im})$ paths, and thus the splitting. We note that this same approach was applied to obtain the tunneling splittings of H-atom and D-atom transfer in malonaldehyde.\cite{qim2008} The results were within roughly 10 percent of the rigorous diffusion Monte Carlo splittings. 

Using this 1d approach, the ground-state tunneling splittings are 37.9 and 8.2 cm$^{-1}$ for H and D, respectively; the H-splitting is in reasonably good agreement with the 32 cm$^{-1}$ obtained from the DMC calculation, and the D-splitting is also consistent with the fact that it is smaller than the uncertainty of the DMC calculations.


It is clearly seen that the $\Delta$-ML approach can indeed bring the PES closer to the coupled cluster level of accuracy, especially as applied to the barrier height of the H-atom transfer. The improvements in geometries and harmonic frequencies are relatively smaller, as the MP2 results are already quite close to the new LCCSD(T) ones presented here.

Of course these improvements are not achieved without extra cost. First, there is an extra cost to compute the LCCSD(T)-F12 energies for 2151 configurations. A single-point calculation of the energy takes about 30 minutes using 12 cores of the 2.4 GHz Intel Xeon processor, and all the 2151 points can be completed within one week using 7 nodes for these computations. As we have shown above, far fewer points can be used to obtain a high-quality correction PES. So, this cost is affordable and minor.

Perhaps of more importance is the extra cost in evaluating the corrected PES. In the $\Delta$-ML approach, the extra cost is the calculation of the energy correction, $\Delta V_{CC-LL}$. For AcAc, the $\Delta{V_{CC-LL}}$ PES uses maximum polynomial order of 2 and it costs about 10\% of of the $V_{LL}$ PES. So this additional cost is also a small price to pay for bringing the accuracy of the PES, especially the barrier height for the H-atom transfer, to near the coupled-cluster level.

The  success in bringing an MP2-based full-dimensional PES for 15-atom acetyacetone to the coupled-cluster quality is very encouraging. Given that a small number (around 500 to 2000) of coupled-cluster energies were needed for the correction makes it clear that the approach should be readily applicable to molecules with more than 10 atoms.  The coupled-cluster approach used was the relatively efficient local method LCCSD(T) available in the 2015 version of Molpro we use.\cite{MOLPRO_brief} Other efficient CCSD(T) methods are also available both in Molpro and other software packages. 

Overall, the success of the current application of the $\Delta$-ML method to 15-atom AcAc and the applications reported earlier on smaller molecules and application to $cis$ and $trans$ $N$-methyl acetamide suggests that the method can have wide applicability and ease of use.  Since the correction PES is not localized around a minimum or a reaction path, it can be used in general anharmonic vibrational analyses of polyatomic molecules. The present application to full dimensional DMC calculations is already evidence of this.

Finally, it is also worth commenting on the current $\Delta$-ML approach and the recent application of transfer learning (TL) by Meuwly and co-workers\cite{meuwly20} to MP2-based PESs for AcAc.  The MP2-based PES we reported used a slightly extended database of MP2 energies and gradients from that group.  Thus, the PES we reported is not the same as the earlier one. However, they are similar, e.g., the H-atom transfer barrier heights are 2.13 and 2.17 kcal/mol for the PIP PES and the neural-network (NN) PES, respectively.  These are in very good agreement with the direct MP2 result of 2.18 kcal/mol.  

Several TL NN models, based on random training data sets (PNO-LCCSD(T) energies) of different sizes, were considered by Meuwly and co-workers.\cite{meuwly20}  In Table 3 of that paper, results both from a single TL NN model and from an average of several TL NN models were given at the optimized geometries of each model. A TL-NN model gave a barrier height of 0.92 kcal/mole using 100 high-level energies, 2.4 kcal/mol from a single TL-NN model, 1.80 kcal/mol averaged from several TL-NN models with 1000 energies, and 2.66 kcal/mol from a single TL-NN model using 5000 energies.  Barrier heights of 3.31 and 3.32 kcal/mol were obtained with a single and multiple TL-NN models using 15 000 energies; these are in excellent agreement with the benchmark barrier height of 3.25 kcal/mol. From these results two apparent conclusions emerge.  The first is that the TL-NN may produce a worse result, i.e., a lower barrier height, than the original NN PES, and the second is that roughly 15 000 high-level energies are needed to obtain a TL-NN PES with an accurate barrier height.  (These authors did note some improvement in the TL-NN model using 1000 energies by the addition of 100 energies along the minimum energy path, i.e., the barrier height went from 1.8 to 2.72 kcal/mol.)

Thus based on the above, it appears the $\Delta$-ML approach performs well and is a reasonable alternative to TL for this example. However, unlike the TL-NN approach, the $\Delta$-ML approach we applied does not produce worse results than the low-level PES.  We attribute this to the fact that the difference potential is small relative to the low-level and high-level potentials.  If this is not true (and this can be checked of course) then the current approach would probably require a larger database of high-level energies to achieve a satisfactory result. These are of course early-days conclusions and more work on both approaches is clearly warranted.

\section*{Acknowledgment}
JMB thanks the ARO, DURIP grant (W911NF-14-1-0471), for funding a computer cluster where most of the calculations were performed and current financial support from NASA (80NSSC20K0360). We are thankful for correspondence with Markus Meuwly and Silvan Käser.

\section*{Supporting Information available}
- Atom numbering scheme;\\
- Table of benchmark harmonic frequencies;\\
- Fitting precision plots.

\bibliography{refs}





\end{document}